# Title: Quantum state resolution of the $C_{60}$ fullerene

**Authors:** P. Bryan Changala[1*], Marissa L. Weichman[1], Kevin F. Lee[2], Martin E. Fermann[2], Jun Ye[1*]

**Affiliations:**

[1]JILA, National Institute of Standards and Technology and University of Colorado, Department of Physics, University of Colorado, Boulder, Colorado 80309, USA.

[2]IMRA America, Inc., 1044 Woodridge Ave., Ann Arbor, Michigan 48105, USA.

*Corresponding authors: bryan.changala@colorado.edu (P.B.C.), ye@jila.colorado.edu (J.Y.)

**Abstract:** The remarkable physical properties of buckminsterfullerene, $C_{60}$, have attracted intense research activity since its original discovery. Total quantum state resolved measurements of isolated $C_{60}$ molecules have been of particularly long-standing interest. However, such observations have to date been unsuccessful due to the difficulty in preparing cold, gas-phase $C_{60}$ in sufficiently high densities. Here we report high resolution infrared absorption spectroscopy of $C_{60}$ in the 8.5 μm spectral region. A combination of cryogenic buffer gas cooling and cavity-enhanced direct frequency comb spectroscopy has enabled the observation of quantum state resolved rovibrational transitions. Characteristic nuclear spin statistical intensity patterns provide striking confirmation of the indistinguishability of the sixty $^{12}C$ atoms, while rovibrational fine structure encodes further details of the molecule's rare icosahedral symmetry. These observations establish new possibilities in the study and control of emergent complexity in finite-sized quantum systems such as fullerenes.

**One Sentence Summary:** High resolution infrared spectroscopy of cold, gas-phase $C_{60}$ reveals fundamental details of its quantum mechanical structure.

**Main Text:**

Understanding molecules as quantum mechanical systems is a central objective of chemical and molecular physics. The complex internal dynamics of these systems evolve over wide energy and time scales as exhibited by the various electronic, vibrational, rotational, and spin degrees of freedom. Polyatomic molecules, in particular, offer the prospect of probing many-body physics in strongly interacting systems. The most comprehensive characterization of a molecular Hamiltonian, which governs intramolecular dynamics, is provided with high resolution spectroscopy. When a polyatomic molecule is sufficiently cold to concentrate the population into and thereby spectrally probe a single rovibrational state, we achieve the unimolecular equivalent of a pure quantum state at absolute zero in the rest frame of the molecule. The precise measurement of transition energies between individual molecular eigenstates yields detailed information about strong, multi-body interactions between atoms in a unimolecular polyatomic lattice, thus providing profound insights into complex molecular structure and ensuing interaction dynamics.

Here we report the first rotationally resolved spectrum of buckminsterfullerene, $C_{60}$. Following the discovery of $C_{60}$ by Kroto *et al.* in 1985 (*1*), infrared and $^{13}C$ NMR spectroscopy confirmed its caged, icosahedral structure (*2-7*). Subsequent spectroscopic and analytical techniques, including x-ray and electron diffraction (*8, 9*), optical Raman and neutron scattering (*10-15*), matrix isolation infrared spectroscopy (see (*16-18*) and references therein), and photoelectron spectroscopy (*19, 20*), have greatly advanced our understanding of this remarkable molecule. Spectroscopy has also played a central role in the astronomical detection of $C_{60}$ and its derivatives (*21, 22*). However, to date, there have been no reports of total quantum state resolved measurements of isolated gas-phase $C_{60}$ molecules. The experiments reported here thus establish $C_{60}$ as the largest molecule, and the only example of rare icosahedral symmetry, for which a complete internal quantum state resolved spectrum has been observed.

While quantum state resolved spectroscopy is routine for small, light molecules, systems as large and heavy as $C_{60}$ are much less amenable to high resolution characterization. Rovibrational spectroscopy in the infrared region is made especially difficult due to several intrinsic and technical challenges. The increase in both the number of vibrational modes and the magnitude of the moment of inertia for every additional atom results in significantly more rotation-vibration states populated at a given internal temperature. Rovibrational states excited by an infrared photon may be strongly coupled to a highly congested manifold of background dark states, the density of which grows rapidly with increasing internal energy, leading to intramolecular vibrational redistribution (*23*). The Doppler broadening of optical transitions due to finite translational temperature serves only to exacerbate this spectral congestion. Furthermore, the low gas-phase densities achievable for heavy, non-volatile species require high detection sensitivity.

These various experimental challenges are addressed by cooling the translational and internal temperatures of gas-phase molecular samples. The method of cryogenic buffer gas cooling is particularly effective for large, heavy molecules (*24, 25*). We have recently demonstrated the integration of a buffer gas cooling source with cavity-enhanced direct frequency comb spectroscopy (CE-DFCS) in the mid-infrared (*26, 27*), which enables sensitive, broadband, high resolution absorption measurements (*28, 29*). We have since extended this apparatus to the long-wave infrared (LWIR) region (*30*) and made significant changes to the buffer gas cooling conditions to permit the preparation and detection of cold, gas-phase samples of even heavier molecules.

Figure 1A depicts a simplified view of the apparatus used for $C_{60}$ cooling and spectroscopy. A 950 K copper oven sublimates solid $C_{60}$ samples, generating gas-phase $C_{60}$ with an average internal energy of 6-8 eV per molecule and occupying up to an estimated $10^{26}$-$10^{30}$ vibrational quantum states, as shown in Fig. 1B. These hot molecules flow into a cell anchored to a cryogenic cold finger, where they are thermalized close to the cell wall temperature via collisions with cold buffer gas atoms introduced through an annular slit inlet plate surrounding the cell entrance aperture. We interrogate the cold gas-phase molecules with CE-DFCS by coupling a frequency comb into a high finesse optical cavity surrounding the cold cell, which enhances the absorption signal by a factor on the order of the cavity finesse ($F = 6000$). The LWIR frequency comb light centered near 8.5 µm is produced by difference frequency generation (DFG) with two near-infrared frequency combs originating from a single mode-locked fiber laser (*31*). The comb contains narrow "teeth" at optical frequencies $v_m = m \times f_{rep} + f_0$, where $f_{rep}$ is the repetition rate. While the DFG comb does not have an offset frequency $f_0$, it can be introduced via an external acousto-optic modulator before the difference frequency step. The intensity of each comb tooth transmitted through the cavity is read out using a broadband scanning-arm Fourier transform interferometer (*32, 33*). Additional experimental details are provided in Materials and Methods (*34*).

Our first attempts at observing cold gas-phase $C_{60}$ with low pressure helium buffer gas conditions similar to our previous work (*26, 27*) yielded no detectable absorption. However, when the vacuum chamber was flooded with a high pressure of helium buffer gas, a single broad, unresolved absorption feature appeared, as shown by the red trace in Fig. 2A. We attribute this spectrum to partially cooled $C_{60}$ molecules that remain warm enough to occupy many vibrational quantum states. This is not surprising: as can be seen in Fig. 1B, even at room temperature the vibrational partition function is greater than $10^3$. This finding suggested that both a higher number of collisions and more efficient energy transfer per collision would be required to thermalize $C_{60}$ to its ground vibrational state (*35*). Indeed, we ultimately produced a sufficiently dense, cold $C_{60}$ sample by (*i*) increasing the buffer gas mass by switching from helium to argon and (*ii*) carefully optimizing the buffer gas flow and oven positioning relative to the inlet slit. The spectrum acquired at these conditions is shown by the blue trace in Fig. 2A and exhibits well resolved rovibrational fine structure, with narrow linewidths on the order of 20 MHz (Fig. S1). The peak absorption, near the band origin, is 10% of the cavity-transmitted comb mode intensity. From the magnitude of the integrated absorption cross-section (*17*), we estimate the number density of cold $C_{60}$ to be $4 \times 10^{11}$ cm$^{-3}$. Observing the appearance and evolution between the broad and narrow signals was greatly facilitated by the wide spectral bandwidth of the frequency comb, which covers the entire breadth of the observed vibrational band. The inferred rotational temperature is approximately 150 K (*34*), nearly equal to the cell wall temperature of 135 K, which is kept well above argon's condensation point of 87 K.

The observed fine structure in the infrared spectrum encodes fundamental details of the quantum mechanical structure of $C_{60}$. To zeroth order, the rotations of $C_{60}$ can be considered as those of a spherical top with total angular momentum operator $\mathbf{J}$ (*36*). The associated rotational quantum states are $|J, k, m\rangle$, where $J = 0, 1, 2, \ldots$ is the total angular momentum quantum number, and $k, m = -J, \ldots, +J$ are the projection quantum numbers of the body-fixed component ($J_z$) and lab-fixed component ($J_Z$) of $\mathbf{J}$, respectively. The triply degenerate vibrational mode of $T_{1u}$ symmetry that gives rise to the infrared band can be modeled as a 3D isotropic harmonic oscillator with vibrational angular momentum operator $\ell$. Its quantum states are $|n, \ell, k_\ell\rangle$, where $n$ is the total number of vibrational quanta, $\ell = n, n-2, n-4, \ldots$ is the vibrational angular momentum quantum number, and $k_\ell = -\ell, \ldots, +\ell$ is the projection quantum number of the body-frame projection ($\ell_z$) of $\ell$.

The uncoupled rovibrational product wavefunctions $|J, k, m\rangle |n, \ell, k_\ell\rangle$ are simultaneously eigenfunctions of $\mathbf{J}^2$, $\ell^2$, $J_z$, $\ell_z$ and $J_Z$. It is useful to define the "pure rotational" angular momentum $\mathbf{R} = \mathbf{J} - \ell$, the eigenfunctions of which can be constructed by transforming the uncoupled product wavefunctions using standard angular momentum coupling relations (*36*). This yields total coupled rovibrational wavefunctions of the form $|R, k_R, J, \ell, n, m\rangle$, where $R$ is the angular momentum quantum number of $\mathbf{R}$ and $k_R = -R, \ldots, +R$ is the body-fixed projection. As usual, the values of $R$ satisfy the triangle inequality $R = |J - \ell|, \ldots, J + \ell$. In this work, we are concerned only with the ground vibrational state with $n = \ell = 0$ and the excited $T_{1u}$ vibrational

state, populated by the IR photon, with $n = \ell = 1$. Therefore, in the ground vibrational state, $R = J$; similarly, in the excited state where $\ell = 1$, $R$ is restricted to $J, |J \pm 1|$.

The energies of the states we observe are determined by effective rotational Hamiltonians for each vibrational state. The simplest possible effective Hamiltonian for the ground vibrational state is that of a rigid spherical top

$$H_{gr} = B'' \mathbf{J}^2, \tag{1}$$

where $B''$ is the ground state rotational constant, which is inversely proportional to the moment of inertia. The ground state wavefunctions $|R = J, k_R, J, \ell = 0, n = 0, m\rangle$ are eigenstates of $H_{gr}$ with energies

$$E_{gr} = B'' J(J+1). \tag{2}$$

This energy is independent of $k_R$ and $m$, leading to the usual $(2R+1)(2J+1) = (2J+1)^2$ spherical top ground state degeneracy factor.

The excited vibrational state is described to lowest order by a slightly more sophisticated effective Hamiltonian,

$$H_{ex} = \nu_0 + B' \mathbf{J}^2 - 2B'\zeta (\mathbf{J} \cdot \boldsymbol{\ell}), \tag{3}$$

where $\nu_0$ is the vibrational band origin, and $B'$ is the excited state rotational constant, which differs slightly from $B''$ due to changes of the moment of inertia upon vibrational excitation. The new rightmost term arises from Coriolis forces that couple the total angular momentum $\mathbf{J}$ and the vibrational angular momentum $\boldsymbol{\ell}$, with $\ell = 1$. The $\zeta$ constant encodes the strength of this coupling, which is determined by the geometric details of the vibrational normal mode. The excited state wavefunctions $|R, k_R, J, \ell = 1, n = 1, m\rangle$ are eigenstates of $H_{ex}$ with energy levels at

$$E_{ex} = \nu_0 + B' J(J+1) - B'\zeta \left[ J(J+1) + \ell(\ell+1) - R(R+1) \right], \tag{4}$$

again with a degeneracy of $(2R+1)(2J+1)$. As $R = J, |J \pm 1|$, the excited state energies sort into three distinct manifolds (*37*)

$$\begin{aligned} E_{ex}^{(+)} &= E_J + 2B'\zeta J, & R &= J+1, \\ E_{ex}^{(0)} &= E_J - 2B'\zeta, & R &= J, \\ E_{ex}^{(-)} &= E_J - 2B'\zeta(J+1), & R &= J-1, \end{aligned} \tag{5}$$

where $E_J = \nu_0 + B' J(J+1)$ is the pure vibrational and rigid rotor contribution to the energy. Physically, these manifolds correspond to states where $\mathbf{J}$ and $\boldsymbol{\ell}$ are mutually antiparallel, perpendicular, and parallel, respectively.

Rovibrational transitions between the ground and excited $T_{1u}$ vibrational states of spherical tops such as $C_{60}$ are governed by the usual strict $\Delta J = 0, \pm 1$ rule and an additional $\Delta R = \Delta k_R = 0$ rule (*36*). These allowed transitions are illustrated in the level diagram of Fig. 2B. Unlike less symmetric molecules, these selection rules dictate that the usual P ($\Delta J = -1$), Q ($\Delta J = 0$), and R ($\Delta J = +1$) transitions reach mutually exclusive sets of upper state quantum levels. These three manifolds are labeled $T_{1u}^{(+)}$, $T_{1u}^{(0)}$, and $T_{1u}^{(-)}$ according to the energy expressions in Eq. (5).

Inspection of the level diagram in Fig. 2B shows that states with certain values of $R$ are missing. This is, in fact, an exceptional example of nuclear spin statistics at work. The carbon nuclei in pure $^{12}C_{60}$ are each identical spin-0 bosons, so any permutation of nuclei must leave the total molecular wavefunction unchanged. This imposes the strict condition that only states with a total rovibronic symmetry of $A_g$ (+ parity) or $A_u$ (− parity) in the $I_h$ point group may exist. Group theoretical analysis (*38*) of the rovibrational wavefunctions shows that this condition is met only with certain linear combinations of $k_R$ states for a given value of $R$. In fact, only a *single* such linear combination is possible for $R = 0, 6, 10, 12, 15, 16, 18, 20$-$22$, and $24$-$28$, with other values of $R < 30$ having no allowed states. (For levels with $R \geq 30$, the number of allowed states is equal to 1 plus the number of states for $R$ minus 30.) The unusual patterns of allowed angular momentum quantum numbers are intimately related to the 2-, 3- and 5-fold symmetry axes of an icosahedron. In the high-$R$ limit, only 1 in 60 states exist due to the drastic effects of these $^{12}C$ nuclear spin statistics.

Taking the zeroth order energies, selection rules, and spin statistics all together, one is left with the predicted spectrum plotted in black in Fig. 2A. It consists of a sharp Q branch surrounded by P and R branches containing lines evenly spaced by approximately $(B'' + B')(1 - \zeta) \approx 0.0078$ cm$^{-1}$. The qualitative appearance of the measured R and Q branch regions is consistent with the simulation, while there is substantial disagreement in the P branch. The portions of the spectrum shown in Fig. 3 provide a closer view of this behavior.

The R branch exhibits a regularly spaced progression of transitions R($J$) that we have assigned from $J \approx 60$-$360$. Transitions outside this range are below our detection sensitivity. Such high values of the total angular momentum quantum number have been rarely observed, if ever, by rotationally resolved frequency domain spectroscopy. Portions of the measured and simulated R branch from $J = 160$-$200$ are shown in Fig. 3A. Despite the noise in the measured absorption, these transitions clearly show the expected discrete intensity variations in the correct integer ratios. Such patterns are a basic consequence of the quantum mechanical indistinguishability and the perfect icosahedral arrangement of the carbon nuclei that make up $^{12}C_{60}$.

Quantitative analysis of the R branch transition frequencies permits extraction of spectroscopic constants. The energy expressions in Eqs. (2) and (5) yield expected transition frequencies of

$$\nu[\text{R}(J)] = \nu_0 + (2\bar{B} + \Delta B)(1 - 2\zeta) \\ + J[2\bar{B}(1-\zeta) + \Delta B(2-\zeta)] \\ + J^2 \Delta B, \quad (6)$$

where $\bar{B} \equiv (B' + B'')/2$ is the mean value of the lower and upper state rotational constants, and $\Delta B \equiv B' - B'' \ll \bar{B}$ is their difference. Figure 4A shows the measured positions (*34*) as a function of lower state $J$, which follow the expected nearly linear dependence. Figure 4B shows the residuals from a fit of Eq. (6) to the measured line positions, displaying two avoided crossings arising from perturbations in the excited state. The fitted spectroscopic parameters are summarized in Table 1. The R branch transition frequencies are well reproduced despite the simplicity of the zeroth order Hamiltonian, which ignores centrifugal distortion effects, and the very high range of *J*. A complete listing of the approximately three hundred transition frequencies used in this fit is given in Data S1.

These measurements represent the first quantum state resolved gas-phase spectrum reported for $C_{60}$. As noted by Brieva *et al.* (*18*) such infrared absorption measurements may help resolve current uncertainties regarding the physical state of astronomical $C_{60}$. More fundamentally, they provide structural information of isolated gas-phase molecules through the rotational fine structure. While the transitions included in our initial analysis do not yet allow an independent determination of $B''$ and $\zeta$, if we assume a range of $\zeta = -0.30$ to $-0.45$ based on theoretical calculations (*37*) we can estimate $B'' = \frac{1}{hc}\frac{\hbar^2}{2I} \approx 0.0027\text{-}0.0030$ cm$^{-1}$, where $I$ is the effective moment of inertia of the ground vibrational state. Given $I = \frac{2}{3}mr^2$ for a spherical shell of mass $m$ and radius $r$, the corresponding range of radii is 3.4 to 3.6 Å. This is consistent with a previous gas-phase electron diffraction measurement of 3.557(5) Å, which includes thermal averaging effects that lengthen the measured radius relative to that of the vibrational ground state (*8*). Furthermore, our measured value of $\Delta B$ implies that the effective $C_{60}$ radius increases by only 0.005% upon excitation of the observed vibrational mode, which is primarily of a surface-tangent C–C bond stretching character. Further analysis of the rotational fine structure of $^{12}C_{60}$ (and ultimately $^{12}C_{59}^{13}C$) will be necessary to constrain $B''$ and $\zeta$ independently and completely determine the gas-phase structural parameters.

The Q branch region is shown in Fig. 3B. There are several unresolved features here, though each is still quite narrow on an absolute scale of 0.01 to 0.03 cm$^{-1}$. The highest frequency feature is assigned as the Q branch of the $^{12}C_{60}$ isotopologue in its ground vibrational state. Centrifugal distortion effects create a band head observed near $J = 250$ (inset of Fig. 3B). The remaining features in the Q branch region are not definitively assigned. While they are possibly hot band transitions of the $^{12}C_{60}$ isotopologue, we believe they most likely derive from the singly substituted $^{12}C_{59}^{13}C$ isotopologue. Despite a $^{13}C$ natural abundance of only 1.1%, the sixty equivalent substitution sites lead to a remarkably high ($^{12}C_{59}^{13}C$):$^{12}C_{60}$ ratio of about 2:3. The substitution breaks the icosahedral symmetry of $C_{60}$, splitting the three-fold degeneracy of the vibrational level and nullifying the nuclear spin statistics. Many more rotational levels and transitions are expected, which will be further split by the non-spherical moments of inertia (*39*).

Finally, two representative portions of the P branch are shown in Fig. 3C. Here, the zeroth order simulation fails to capture either the position or number of observed transitions. This complicated fine structure is likely due to high order centrifugal distortion terms not included in the simulated spectrum (*40*). The zeroth order Hamiltonians, Eqs. (1) and (3), contain only scalar terms that preserve the spherical degeneracy of the (2R+1) body-fixed projections of **R**. While most of these sub-states are eliminated by the $^{12}C$ nuclear spin statistics, the degeneracy of the remaining sub-states can be broken by non-scalar centrifugal distortion terms. These so-called "icosahedral splitting" terms (*40*) lead to subsequent splittings of the observed transitions. In the ground state, the lowest order non-scalar centrifugal distortion term scales like $J^6$, whereas such terms can appear in the excited state that scale only as $J^4$. Due to the large $J$ values observed here, it is not surprising that such effects become important. However, to date there have been no theoretical predictions of the magnitude of these icosahedral splitting terms. A full analysis of this portion of the spectrum is most effectively treated within the irreducible spherical tensor formalism (*41*) and is ongoing in our laboratory.

The present experiments represent the beginning of what we hope is a new avenue of $C_{60}$ research and fullerene science. The general applicability of buffer gas cooling establishes the possibility of similar studies, both in the infrared and other spectral regions, on larger fullerenes such as $C_{70}$; endofullerenes, wherein an atom or small molecule is encapsulated in a closed fullerene cage; or even pure $^{13}C_{60}$, which represents a pristine example of a spin-½ network on a spherical lattice. Ultimately, precision spectroscopy of such targets is the first step towards single quantum state preparation and control of large molecular systems.

**References and Notes:**
1. H. W. Kroto, J. R. Heath, S. C. O'Brien, R. F. Curl, R. E. Smalley, $C_{60}$: Buckminsterfullerene. *Nature* **318**, 162-163 (1985).
2. H. Ajie, M. M. Alvarez, S. J. Anz, R. D. Beck, F. Diederich, K. Fostiropoulos, D. R. Huffman, W. Krätschmer, Y. Rubin, K. E. Schriver, D. Sensharma, R. L. Whetten, Characterization of the Soluble All-Carbon Molecules $C_{60}$ and $C_{70}$. *J. Phys. Chem.* **94**, 8630-8633 (1990).
3. R. Taylor, J. P. Hare, A. K. Abdulsada, H. W. Kroto, Isolation, Separation and Characterization of the Fullerenes $C_{60}$ and $C_{70}$: the Third Form of Carbon. *J. Chem. Soc., Chem. Commun.* **1990**, 1423-1424 (1990).
4. W. Krätschmer, K. Fostiropoulos, D. R. Huffman, The Infrared and Ultraviolet Absorption Spectra of Laboratory-Produced Carbon Dust: Evidence for the Presence of the $C_{60}$ Molecule. *Chem. Phys. Lett.* **170**, 167-170 (1990).
5. W. Krätschmer, L. D. Lamb, K. Fostiropoulos, D. R. Huffman, Solid $C_{60}$: a New Form of Carbon. *Nature* **347**, 354-358 (1990).
6. C. S. Yannoni, P. P. Bernier, D. S. Bethune, G. Meijer, J. R. Salem, NMR Determination of the Bond Lengths in $C_{60}$. *J. Am. Chem. Soc.* **113**, 3190-3192 (1991).
7. C. S. Yannoni, R. D. Johnson, G. Meijer, D. S. Bethune, J. R. Salem, $^{13}C$ NMR Study of the C60 Cluster in the Solid State: Molecular Motion and Carbon Chemical Shift Anisotropy. *J. Phys. Chem.* **95**, 9-10 (1991).
8. K. Hedberg, L. Hedberg, D. S. Bethune, C. A. Brown, H. C. Dorn, R. D. Johnson, M. De Vries, Bond Lengths in Free Molecules of Buckminsterfullerene, $C_{60}$, from Gas-Phase Electron Diffraction. *Science* **254**, 410-412 (1991).
9. S. Liu, Y.-J. Lu, M. M. Kappes, J. A. Ibers, The Structure of the $C_{60}$ Molecule: X-Ray Crystal Structure Determination of a Twin at 110 K. *Science* **254**, 408-410 (1991).
10. S. F. Parker, S. M. Bennington, J. W. Taylor, H. Herman, I. Silverwood, P. Albers, K. Refson, Complete assignment of the vibrational modes of $C_{60}$ by inelastic neutron scattering spectroscopy and periodic-DFT. *Phys. Chem. Chem. Phys.* **13**, 7789-7804 (2011).

11. L. Pintschovius, Neutron studies of vibrations in fullerenes. *Rep. Prog. Phys.* **59**, 473-510 (1996).
12. D. S. Bethune, G. Meijer, W. C. Tang, H. J. Rosen, The Vibrational Raman Spectra of Purified Solid Films of $C_{60}$ and $C_{70}$. *Chem. Phys. Lett.* **174**, 219-222 (1990).
13. D. S. Bethune, G. Meijer, W. C. Tang, H. J. Rosen, W. G. Golden, H. Seki, C. A. Brown, M. S. De Vries, Vibrational Raman and Infrared Spectra of Chromatographically Separated $C_{60}$ and $C_{70}$ Fullerene Clusters. *Chem. Phys. Lett.* **179**, 181-186 (1991).
14. K. Prassides, T. J. S. Dennis, J. P. Hare, J. Tomkinson, H. W. Kroto, R. Taylor, D. R. M. Walton, Inelastic Neutron Scattering Spectrum of the Fullerene $C_{60}$. *Chem. Phys. Lett.* **187**, 455-458 (1991).
15. R. L. Cappelletti, J. R. D. Copley, W. A. Kamitakahara, F. Li, J. S. Lannin, D. Ramage, Neutron Measurements of Intramolecular Vibrational Modes in $C_{60}$. *Phys. Rev. Lett.* **66**, 3261-3264 (1991).
16. N. Sogoshi, Y. Kato, T. Wakabayashi, T. Momose, S. Tam, M. E. DeRose, M. E. Fajardo, High-resolution infrared absorption spectroscopy of $C_{60}$ molecules and clusters in parahydrogen solids. *J. Phys. Chem. A* **104**, 3733-3742 (2000).
17. S. Iglesias-Groth, F. Cataldo, A. Manchado, Infrared spectroscopy and integrated molar absorptivity of $C_{60}$ and $C_{70}$ fullerenes at extreme temperatures. *Mon. Not. R. Astron. Soc.* **413**, 213-222 (2011).
18. A. C. Brieva, R. Gredel, C. Jäger, F. Huisken, T. Henning, $C_{60}$ as a Probe for Astrophysical Environments. *Astrophys. J.* **826**, (2016).
19. X.-B. Wang, C.-F. Ding, L.-S. Wang, High resolution photoelectron spectroscopy of $C_{60}^-$. *J. Chem. Phys.* **110**, 8217-8220 (1999).
20. D.-L. Huang, P. D. Dau, H.-T. Liu, L.-S. Wang, High-resolution photoelectron imaging of cold $C_{60}^-$ anions and accurate determination of the electron affinity of $C_{60}$. *J. Chem. Phys.* **140**, (2014).
21. J. Cami, J. Bernard-Salas, E. Peeters, S. E. Malek, Detection of $C_{60}$ and $C_{70}$ in a Young Planetary Nebula. *Science* **329**, 1180-1182 (2010).
22. E. K. Campbell, M. Holz, D. Gerlich, J. P. Maier, Laboratory confirmation of $C_{60}^+$ as the carrier of two diffuse interstellar bands. *Nature* **523**, 322 (2015).
23. D. J. Nesbitt, R. W. Field, Vibrational energy flow in highly excited molecules: Role of intramolecular vibrational redistribution. *J. Phys. Chem.* **100**, 12735-12756 (1996).
24. D. Patterson, E. Tsikata, J. M. Doyle, Cooling and collisions of large gas phase molecules. *Phys. Chem. Chem. Phys.* **12**, 9736-9741 (2010).
25. J. Piskorski, D. Patterson, S. Eibenberger, J. M. Doyle, Cooling, Spectroscopy and Non-Sticking of *trans*-Stilbene and Nile Red. *Chemphyschem* **15**, 3800-3804 (2014).
26. P. B. Changala, B. Spaun, D. Patterson, J. M. Doyle, J. Ye, Sensitivity and resolution in frequency comb spectroscopy of buffer gas cooled polyatomic molecules. *Appl. Phys. B.* **122**, (2016).
27. B. Spaun, P. B. Changala, D. Patterson, B. J. Bjork, O. H. Heckl, J. M. Doyle, J. Ye, Continuous probing of cold complex molecules with infrared frequency comb spectroscopy. *Nature* **533**, 517 (2016).
28. F. Adler, M. J. Thorpe, K. C. Cossel, J. Ye, Cavity-Enhanced Direct Frequency Comb Spectroscopy: Technology and Applications. *Annu. Rev. Anal. Chem.* **3**, 175-205 (2010).
29. M. J. Thorpe, J. Ye, Cavity-enhanced direct frequency comb spectroscopy. *Appl. Phys. B* **91**, 397-414 (2008).

30. K. Iwakuni, G. Porat, T. Q. Bui, B. J. Bjork, S. B. Schoun, O. H. Heckl, M. E. Fermann, J. Ye, Phase-stabilized 100 mW frequency comb near 10 um. *Appl. Phys. B* **124**, 128 (2018).
31. K. F. Lee, C. J. Hensley, P. G. Schunemann, M. E. Fermann, Midinfrared frequency comb by difference frequency of erbium and thulium fiber lasers in orientation-patterned gallium phosphide. *Opt. Express* **25**, 17411-17416 (2017).
32. J. Mandon, G. Guelachvili, N. Picqué, Fourier transform spectroscopy with a laser frequency comb. *Nat. Photonics* **3**, 99-102 (2009).
33. F. Adler, P. Masłowski, A. Foltynowicz, K. C. Cossel, T. C. Briles, I. Hartl, J. Ye, Mid-infrared Fourier transform spectroscopy with a broadband frequency comb. *Opt. Express* **18**, 21861-21872 (2010).
34. See supplementary material.
35. J. T. Stewart, B. E. Brumfield, B. M. Gibson, B. J. McCall, Inefficient Vibrational Cooling of $C_{60}$ in a Supersonic Expansion. *ISRN Phys. Chem.* **2013**, 1-10 (2013).
36. C. di Lauro, in *Rotational Structure in Molecular Infrared Spectra*. (Elsevier, 2013), chap. 10, pp. 225-245.
37. D. E. Weeks, W. G. Harter, Rotation Vibration Scalar Coupling Zeta Coefficients and Spectroscopic Band Shapes of Buckminsterfullerene. *Chem. Phys. Lett.* **176**, 209-216 (1991).
38. P. R. Bunker, P. Jensen, Spherical top molecules and the molecular symmetry group. *Mol. Phys.* **97**, 255-264 (1999).
39. T. C. Reimer, W. G. Harter, Fullerene symmetry reduction and rotational level fine structure: The buckyball isotopomer $^{12}C_{59}^{13}C$. *J. Chem. Phys.* **106**, 1326-1335 (1997).
40. W. G. Harter, D. E. Weeks, Rotation-Vibration Spectra of Icosahedral Molecules. I. Icosahedral Symmetry Analysis and Fine Structure. *J. Chem. Phys.* **90**, 4727-4743 (1989).
41. V. Boudon, J.-P. Champion, T. Gabard, M. Loëte, F. Michelot, G. Pierre, M. Rotger, C. Wenger, M. Rey, Symmetry-adapted tensorial formalism to model rovibrational and rovibronic spectra of molecules pertaining to various point groups. *J. Mol. Spectrosc.* **228**, 620-634 (2004).
42. J. Tennyson, P. F. Bernath, L. R. Brown, A. Campargue, A. G. Csaszar, L. Daumont, R. R. Gamache, J. T. Hodges, O. V. Naumenko, O. L. Polyansky, L. S. Rothman, A. C. Vandaele, N. F. Zobov, A. R. Al Derzi, C. Fabri, A. Z. Fazliev, T. Furtenbacher, I. E. Gordon, L. Lodi, I. I. Mizus, IUPAC critical evaluation of the rotational-vibrational spectra of water vapor, Part III: Energy levels and transition wavenumbers for $H_2^{16}O$. *J. Quant. Spectrosc. Radiat. Transfer* **117**, 29-58 (2013).
43. W. B. Olson, A. G. Maki, W. J. Lafferty, Tables of $N_2O$ Absorption Lines for the Calibration of Tunable Infrared Lasers from 522 $cm^{-1}$ to 657 $cm^{-1}$ and from 1115 $cm^{-1}$ to 1340 $cm^{-1}$. *J. Phys. Chem. Ref. Data* **10**, 1065-1084 (1981).
44. C. Camy-Peyret, J.-M. Flaud, A. Mahmoudi, G. Guelachvili, J. W. C. Johns, Line Positions and Intensities in the $\nu_2$ Band of $D_2O$ Improved Pumped $D_2O$ Laser Frequencies. *Int. J. Infrared Millimeter Waves* **6**, 199-233 (1985).

**Acknowledgments:** The authors thank Hans Green for technical advice during the design of the oven source and John Doyle for insightful discussions. **Funding:** This work was supported by AFOSR Grant No. FA9550-15-1-0111, the DARPA SCOUT Program, NIST, and NSF PHYS-1734006. M.L.W. is supported through an NRC Postdoctoral Fellowship; **Author contributions:** P.B.C., M.L.W., and J.Y. performed the experiment. K.F.L. and M.E.F. built the difference frequency generation-based comb. All authors contributed to the writing of this paper; **Competing **interests:** We declare no competing interests. **Data and materials availability:** All data is available in the main text or the supplementary materials.

**Supplementary Materials:**

Materials and Methods

Figure S1

Data S1

References (*42-44*)

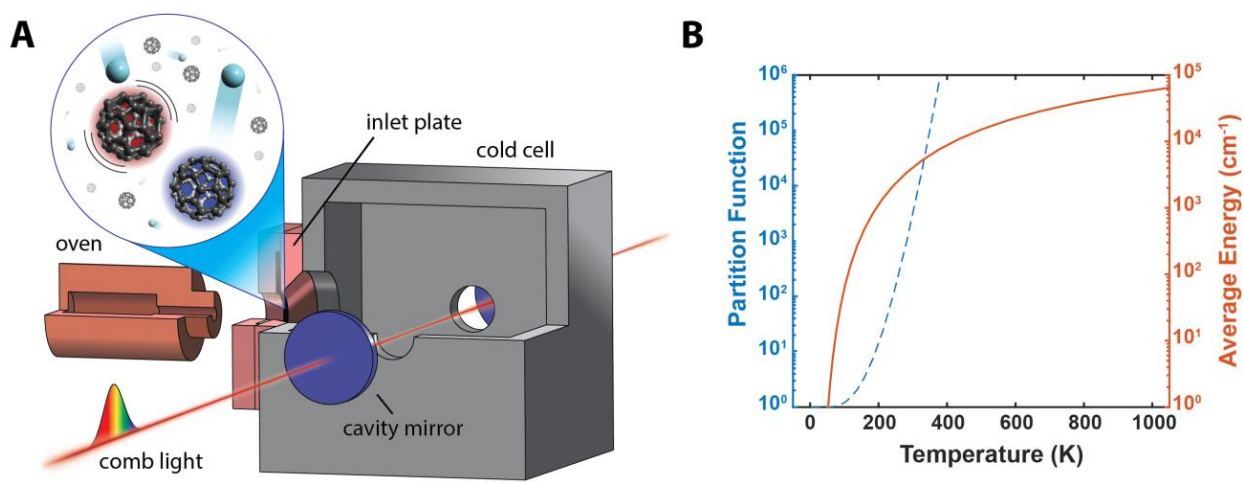

**Fig. 1: Cooling and comb spectroscopy of gas-phase $C_{60}$.** (A) Sublimated $C_{60}$ vapor exits the oven source and enters a cryogenic cell, where it thermalizes via collisions with cold buffer gas introduced through an annular slit inlet plate surrounding the entrance aperture (see inset). Mid-infrared frequency comb light is coupled to an optical enhancement cavity surrounding the cell. The optical absorption spectrum is measured with a scanning arm Fourier transform spectrometer (not pictured). (B) The vibrational partition function (blue dashed line) and average vibrational energy (red solid line) increase strongly as a function of temperature. Approximately 6-8 eV of vibrational energy must be removed per molecule to cool $C_{60}$ from the initial oven temperature to below 150 K, at which point the vibrational partition function is approximately equal to unity.

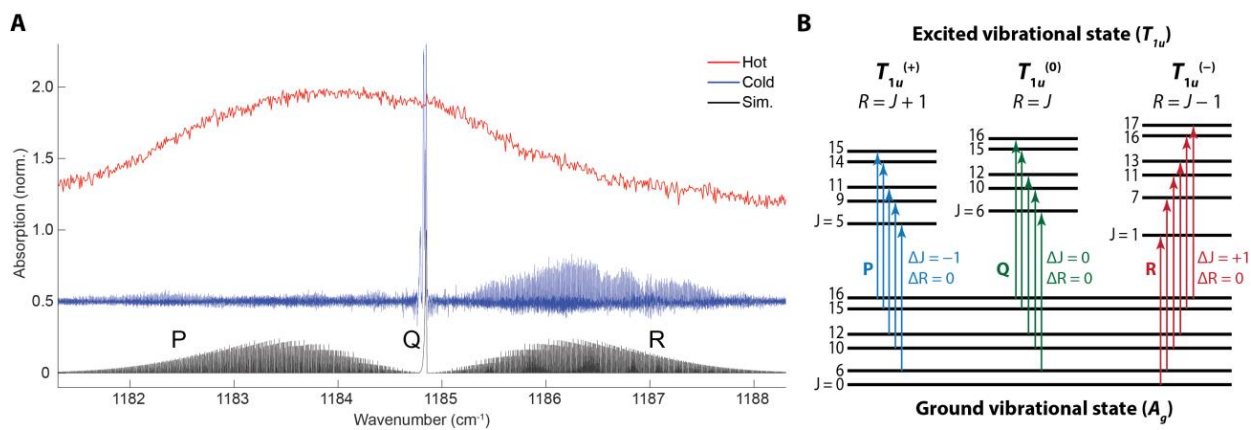

**Fig. 2: Spectroscopic patterns of the IR active vibrational band of $^{12}C_{60}$ near 8.5 μm.** (A) A simulated spectrum (bottom, black) is compared to a measured spectrum of cold (middle, blue) and hot (top, red) $C_{60}$. The measured hot spectrum shows broad, unresolved absorption due to many thermally occupied vibrational states. The cold spectrum exhibits sharp, well resolved rotational structure from transitions out of the ground vibrational state. (B) Rovibrational transitions between the ground vibrational state and the excited state follow zeroth order selection rules of $\Delta J = 0, \pm 1$ and $\Delta R = 0$. These lead to independent P ($\Delta J = -1$), Q ($\Delta J = 0$) and R ($\Delta J = +1$) branches that access the upper state manifolds labeled $T_{1u}^{(+)}$ (for $R = J + 1$), $T_{1u}^{(0)}$ (for $R = J$), and $T_{1u}^{(-)}$ (for $R = J - 1$), respectively.

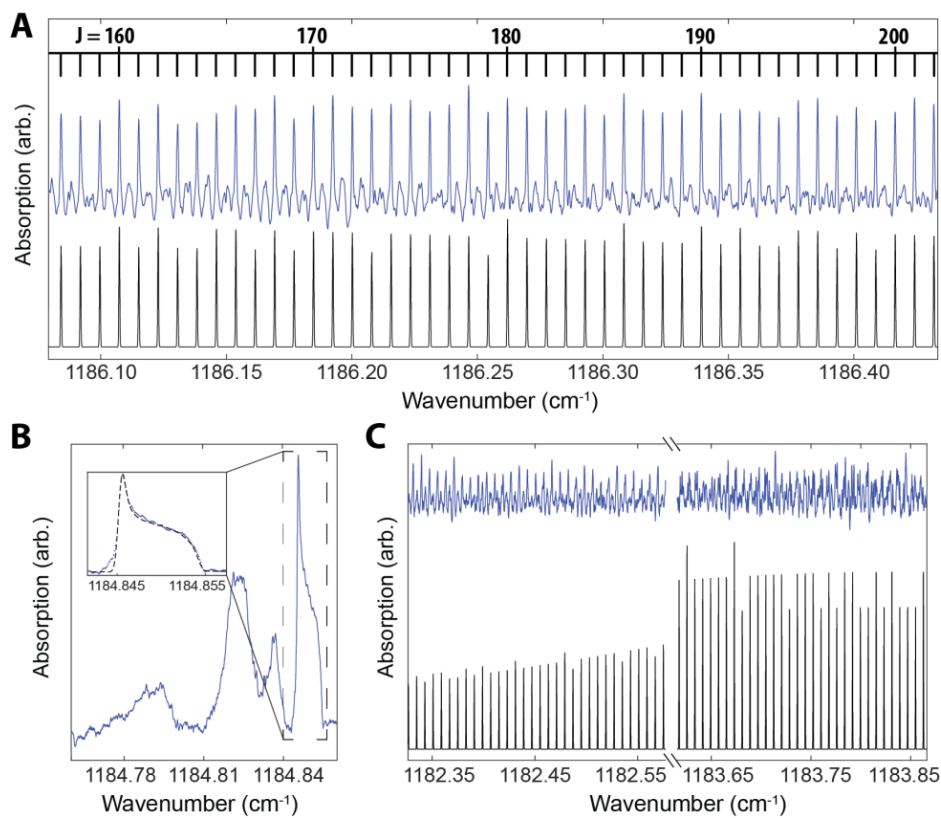

**Fig. 3: Detailed views of portions of the measured IR band.** (A) The R branch shows agreement between the expected intensity patterns from the simulation (bottom, black) and the measured spectrum (top, blue). The tie line above the spectrum indicates the lower state $J$ value of each observed R($J$) transition. (B) The Q branch region of the spectrum contains several features. The highest wavenumber feature is assigned as the Q branch of the $^{12}C_{60}$ isotopologue. Inset: the dashed line represents a fit to a simple quartic centrifugal distortion contour. The additional features at lower frequencies are likely due to the singly substituted $^{13}C^{12}C_{59}$ isotopologue. (C) These two portions of the P branch (top, blue) are representative of the disagreement with the zeroth order simulation determined from parameters fitted to the R branch (bottom, black). The structure not captured by the simulation is evidence of non-scalar centrifugal distortion effects.

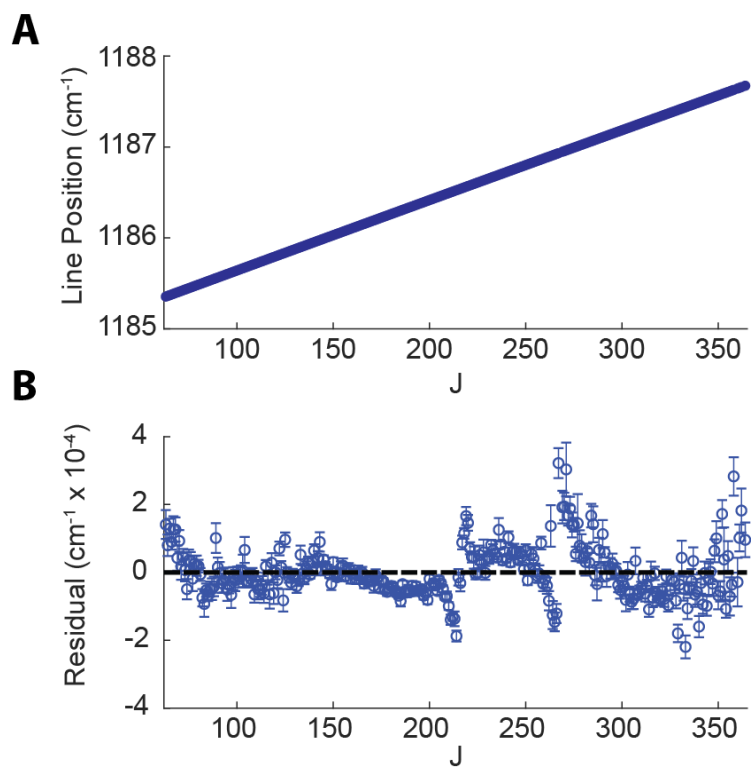

**Fig. 4: Fit results for the R branch.** (A) The R(*J*) line positions plotted versus lower state *J* display a very linear trend over *J* = 60-360. The individual line positions are listed in Data S1. (B) The residuals from the fit of Eq. (6) to these line positions, summarized in Table 1, exhibit apparent avoided crossings near *J* = 215 and *J* = 275, which are possible signatures of local dark state perturbers in the upper state.

**Table 1: Fitted spectroscopic parameters of Eq. (6) for the R branch.** The residuals (Fig. 4B) have a small root-mean-square error of $7.4 \times 10^{-5}$ cm$^{-1}$, slightly larger than the 1σ line-center measurement uncertainty of $2.5 \times 10^{-5}$ cm$^{-1}$.

| Parameter | Value (cm$^{-1}$) |
|---|---|
| $\nu_0 + (2\bar{B} + \Delta B)(1 - 2\zeta)$ | 1184.86196(3) |
| $2\bar{B}(1-\zeta) + \Delta B(2-\zeta)$ | 0.0078300(3) |
| $\Delta B$ | $-2.876(6) \times 10^{-7}$ |

# Supplementary Materials for

**Quantum state resolution of the $C_{60}$ fullerene**

P. Bryan Changala, Marissa L. Weichman, Kevin F. Lee, Martin E. Fermann, Jun Ye

Correspondence to: bryan.changala@colorado.edu (P.B.C.), ye@jila.colorado.edu (J.Y.)

**This PDF file includes:**

    Materials and Methods
    Fig. S1
    Data S1

**Materials and Methods**

Experimental details

$C_{60}$ is vaporized in a small cylindrical copper oven approximately 5 cm long and 3 cm in diameter, loaded with a few hundred milligrams of solid $C_{60}$ powder (Fig. 1A, main text). The oven is heated to 950 K with two silicon nitride heaters clamped to the oven body. Gas exits the oven through a 3 mm hole at its end and enters an aluminum cold cell, $6 \times 6 \times 6$ cm$^3$ in volume, anchored to a liquid nitrogen cold finger in vacuum. An annular slit buffer gas inlet plate surrounds the entrance aperture of the cold cell. At optimum conditions, a flow of 50-100 sccm of argon buffer gas is introduced into the cell through the annular inlet and the background chamber pressure is kept at 250 mTorr. The oven is placed 0.5 to 1.0 cm from the inlet aperture.

LWIR DFG comb light is generated with near-IR driving combs in an orientation patterned GaP crystal (*31*). The two driving fields originate from the same mode-locked erbium-doped fiber oscillator and therefore have an identical repetition rate $f_{rep}$ = 93.4 MHz and carrier-envelope phase offset frequencies, $f_0$. The resultant DFG light has comb teeth at frequencies equal to the difference of the two driving combs. Therefore, it has a passively stable $f_0 = 0$. We generate a non-zero DFG $f_0$ by inserting an acousto-optic modulator in one of the driving beam arms.

An absorption enhancement cavity is formed with two high reflectivity, low loss mirrors that surround the cell. They are spaced by 53.5 cm resulting in a cavity FSR of 280 MHz. Approximately 1 mW of comb light spanning 1150-1200 cm$^{-1}$ is resonantly coupled to the cavity. As the cavity FSR = $3 \times f_{rep}$, only one out of every three comb modes is coupled into the cavity. Frequency sidebands are applied to the light at 2.65 MHz via an electro-optic modulator (EOM) in the oscillator cavity and are used to generate a Pound-Drever-Hall comb-cavity error signal. This error signal is filtered, amplified, and fed back to the EOM and to piezoelectric elements in the laser oscillator cavity to maintain the FSR = $3 \times f_{rep}$ resonance condition. An oscillator monitor photodiode is used to measure the 11$^{th}$ harmonic of $f_{rep}$, which is phase-locked to a local oscillator by feeding back on enhancement cavity length piezos, simultaneously stabilizing the absolute values of $f_{rep}$ and the cavity FSR.

When locked, 20-30% of the resonant comb mode power is transmitted through the cavity. The spectrum of the cavity transmission is measured with a scanning-arm Fourier transform interferometer (*32,33*). The LWIR interferogram is sampled at the zero-crossings of a 1 μm cw-Nd:YAG NPRO laser that simultaneously propagates through the interferometer. Both output ports of the LWIR beam splitter are used to obtain a difference interferogram with an autobalanced circuit, suppressing common mode intensity noise. The instrument resolution of the interferometer (100 MHz) is sufficiently narrow to resolve individual comb teeth, spaced by the FSR of the cavity (280 MHz). We fill the FSR "gap" between cavity resonances by scanning the FSR and interleaving these spectra.

The cold spectrum plotted in blue in Fig 2A in the main text contains data averaged from 78 FSR scans, each consisting of 89 FSR positions evenly distributed over the 280 MHz span. This represents approximately 20 hours of total acquisition time

Frequency calibration and spectral fitting

A preliminary calibration of the optical frequencies is performed by exploiting absorption lines in the spectrum from trace $H_2O$ inside the optical cavity. We first scale the nominal wavenumber axis from the interferometer trace for a given FSR to match known $H_2O$ line positions in the 8.5 µm region (*42*). This provides sufficient accuracy to determine the integer mode index $m$ of each comb tooth, which then determines the final absolute optical frequency via $\nu_m = m \times f_{rep} + f_0$. The values of $f_{rep}$ and $f_0$ are ultimately locked to an atomic time standard with a relative frequency offset error below $10^{-12}$. The absolute frequencies were verified by checking the recalibrated trace $H_2O$ line positions, as well as $N_2O$ and $D_2O$ line positions obtained in separate measurements. All were consistent with reference literature values (*42-44*).

The R branch transitions were fit to Gaussian profiles to determine the linewidth, intensity, and center position of each transition. These parameters are summarized in Data S1. An example fit of the R(180) transition is shown in Fig. S1A. The absolute line center measurement uncertainty is $2.5 \times 10^{-5}$ cm$^{-1}$, dominated by the statistical uncertainty given by our measurement signal-to-noise and the absorption line widths, which range from $0.5 \times 10^{-3}$ cm$^{-1}$ to $1.0 \times 10^{-3}$ cm$^{-1}$. The intensities of the R branch transitions permit extraction of the relative rotational populations in the ground vibrational state, which display a Boltzmann distribution with an effective rotational temperature $T_{rot} = 150(2)$ K (Fig. S1B).

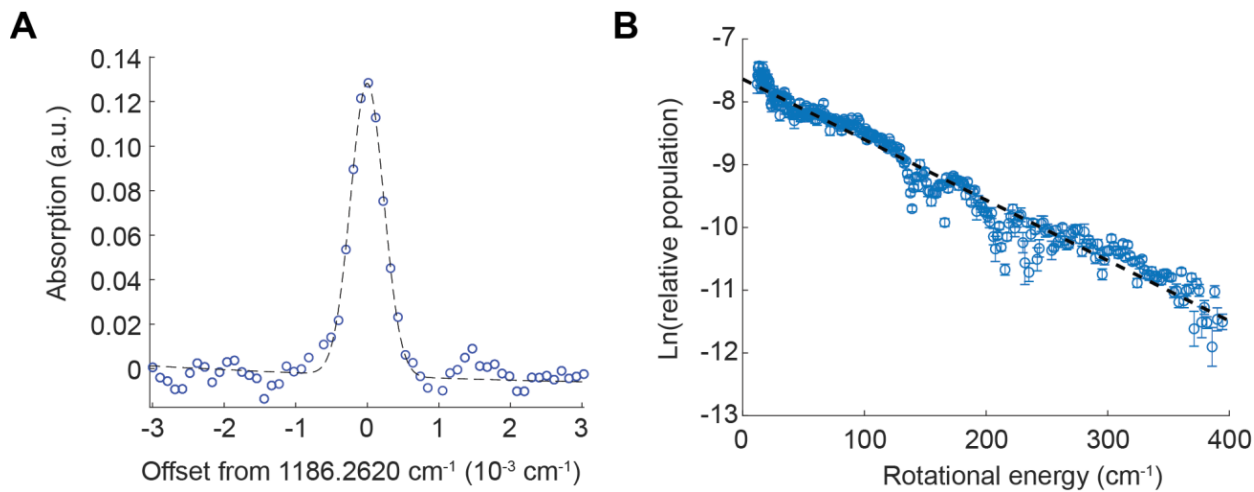

**Fig. S1. Absorption line shape and temperature fits.** (A) The absorption line assigned to the R(180) transition (blue circles) is fitted to a Gaussian profile (black dashed line) with center position 1186.261999(8) cm$^{-1}$ and full-width-at-half-maximum (FWHM) of 16.0(6) MHz. (B) The natural logarithms of the relative ground state rotational level populations (blue circles) are plotted against the ground state rotational energy. The linear trend indicates a Boltzmann distribution with rotational temperature $T_{rot}$ = 150(2) K (black dashed line).

**Data S1 (separate file). Fitted parameters for individual R branch transitions.** This table summarizes the least-squares fitted line center frequency, full-width-at-half-maximum (FWHM), and relative intensity of R(J) transitions for J = 63-364. The mean line center uncertainty is 2.5 × 10$^{-5}$ cm$^{-1}$.

**Data S1.** Fitted parameters for individual R branch transitions. This table summarizes the least-squares fitted line center frequency, full-width-at-half-maximum (FWHM), and relative intensity of R(J) transitions for J = 63-364. The mean line center uncertainty is $2.5 \times 10^{-5}$ cm$^{-1}$.

| Branch | J | Fitted Line Center (cm$^{-1}$) | Uncertainty | FHWM (MHz) | Unc. | Intensity (a.u) | Unc. | Comment |
|---|---|---|---|---|---|---|---|---|
| R | 63 | 1185.35424585 | 0.00004268 | 19.14 | 2.95 | 0.0288 | 0.0043 | |
| R | 64 | 1185.36197804 | 0.00003169 | 22.82 | 2.24 | 0.0334 | 0.0028 | |
| R | 65 | 1185.36982044 | 0.00002900 | 21.28 | 2.05 | 0.0384 | 0.0032 | |
| R | 66 | 1185.37757323 | 0.00002253 | 21.07 | 1.59 | 0.0519 | 0.0034 | |
| R | 67 | 1185.38540278 | 0.00003757 | 33.99 | 2.65 | 0.0407 | 0.0027 | |
| R | 68 | 1185.39319309 | 0.00003552 | 19.95 | 2.51 | 0.0339 | 0.0037 | |
| R | 69 | 1185.40089960 | 0.00003726 | 20.68 | 2.42 | 0.0342 | 0.0037 | |
| R | 70 | 1185.40873990 | 0.00003218 | 23.04 | 2.27 | 0.0448 | 0.0038 | |
| R | 71 | 1185.41645406 | 0.00003947 | 29.48 | 2.79 | 0.0386 | 0.0032 | |
| R | 72 | 1185.42427124 | 0.00002310 | 22.95 | 1.63 | 0.0565 | 0.0035 | |
| R | 73 | 1185.43202051 | 0.00003286 | 12.98 | 2.32 | 0.0368 | 0.0057 | |
| R | 74 | 1185.43975187 | 0.00002988 | 19.43 | 2.11 | 0.0435 | 0.0041 | |
| R | 75 | 1185.44762428 | 0.00003147 | 25.20 | 2.10 | 0.0490 | 0.0037 | |
| R | 76 | 1185.45540796 | 0.00002710 | 25.16 | 1.91 | 0.0503 | 0.0033 | |
| R | 77 | 1185.46314285 | 0.00003698 | 24.06 | 2.61 | 0.0400 | 0.0038 | |
| R | 78 | 1185.47097227 | 0.00002255 | 20.29 | 1.59 | 0.0624 | 0.0042 | |
| R | 79 | 1185.47875112 | 0.00002965 | 19.23 | 2.09 | 0.0402 | 0.0038 | |
| R | 80 | 1185.48653437 | 0.00001873 | 20.91 | 1.32 | 0.0583 | 0.0032 | |
| R | 81 | 1185.49429358 | 0.00001909 | 21.74 | 1.27 | 0.0582 | 0.0029 | |
| R | 82 | 1185.50200781 | 0.00002540 | 29.02 | 1.79 | 0.0600 | 0.0032 | |
| R | 83 | 1185.50977212 | 0.00003768 | 28.19 | 2.66 | 0.0406 | 0.0033 | |
| R | 84 | 1185.51759455 | 0.00001952 | 19.92 | 1.38 | 0.0584 | 0.0035 | |
| R | 85 | 1185.52536049 | 0.00001683 | 20.12 | 1.19 | 0.0598 | 0.0031 | |
| R | 86 | 1185.53315998 | 0.00002239 | 22.89 | 1.58 | 0.0546 | 0.0033 | |
| R | 87 | 1185.54094013 | 0.00003179 | 28.74 | 2.14 | 0.0518 | 0.0033 | |
| R | 88 | 1185.54872794 | 0.00002867 | 27.97 | 2.02 | 0.0436 | 0.0027 | |
| R | 89 | 1185.55664970 | 0.00004396 | 29.24 | 3.10 | 0.0284 | 0.0026 | |
| R | 90 | 1185.56434195 | 0.00002859 | 18.73 | 2.02 | 0.0652 | 0.0061 | |
| R | 91 | 1185.57210311 | 0.00003231 | 25.00 | 2.28 | 0.0530 | 0.0042 | |
| R | 92 | 1185.57985054 | 0.00002848 | 25.09 | 2.01 | 0.0564 | 0.0039 | |
| R | 93 | 1185.58763468 | 0.00003497 | 29.44 | 2.39 | 0.0512 | 0.0035 | |
| R | 94 | 1185.59541831 | 0.00003099 | 24.71 | 2.19 | 0.0489 | 0.0037 | |
| R | 95 | 1185.60315991 | 0.00003314 | 30.45 | 2.34 | 0.0541 | 0.0036 | |
| R | 96 | 1185.61094782 | 0.00002324 | 31.04 | 1.64 | 0.0736 | 0.0034 | |
| R | 97 | 1185.61869410 | 0.00003831 | 30.82 | 2.70 | 0.0470 | 0.0036 | |
| R | 98 | 1185.62651645 | 0.00003343 | 21.25 | 2.36 | 0.0482 | 0.0046 | |
| R | 99 | 1185.63428416 | 0.00003167 | 30.41 | 2.22 | 0.0513 | 0.0031 | |
| R | 100 | 1185.64207236 | 0.00002515 | 26.03 | 1.78 | 0.0680 | 0.0040 | |
| R | 101 | 1185.64983920 | 0.00003020 | 21.19 | 2.13 | 0.0416 | 0.0036 | |
| R | 102 | 1185.65762641 | 0.00001848 | 18.85 | 1.30 | 0.0747 | 0.0045 | |
| R | 103 | 1185.66539970 | 0.00004810 | 38.65 | 3.40 | 0.0579 | 0.0044 | |

| | | | | | | | |
|---|---|---|---|---|---|---|---|
| R | 104 | 1185.67323207 | 0.00003855 | 31.42 | 2.76 | 0.0547 | 0.0041 |
| R | 105 | 1185.68090914 | 0.00001679 | 23.63 | 1.22 | 0.0791 | 0.0034 |
| R | 106 | 1185.68867766 | 0.00001698 | 20.58 | 1.20 | 0.0792 | 0.0040 |
| R | 107 | 1185.69645552 | 0.00002512 | 31.35 | 1.77 | 0.0638 | 0.0031 |
| R | 108 | 1185.70423612 | 0.00001867 | 13.90 | 1.32 | 0.0655 | 0.0054 |
| R | 109 | 1185.71194481 | 0.00002460 | 16.39 | 1.74 | 0.0565 | 0.0052 |
| R | 110 | 1185.71975049 | 0.00002559 | 20.56 | 1.82 | 0.0790 | 0.0060 |
| R | 111 | 1185.72749293 | 0.00002915 | 17.69 | 2.06 | 0.0681 | 0.0069 |
| R | 112 | 1185.73525949 | 0.00002280 | 17.00 | 1.61 | 0.0642 | 0.0053 |
| R | 113 | 1185.74303598 | 0.00002731 | 32.10 | 1.93 | 0.0592 | 0.0031 |
| R | 114 | 1185.75077680 | 0.00002752 | 29.95 | 1.94 | 0.0762 | 0.0043 |
| R | 115 | 1185.75862079 | 0.00001891 | 22.01 | 1.34 | 0.0726 | 0.0038 |
| R | 116 | 1185.76635157 | 0.00002079 | 17.36 | 1.49 | 0.0676 | 0.0050 |
| R | 117 | 1185.77406457 | 0.00004171 | 34.35 | 2.99 | 0.0652 | 0.0048 |
| R | 118 | 1185.78192048 | 0.00002797 | 22.66 | 1.97 | 0.0651 | 0.0049 |
| R | 119 | 1185.78959260 | 0.00003175 | 15.78 | 2.24 | 0.0449 | 0.0055 |
| R | 120 | 1185.79741540 | 0.00002041 | 22.30 | 1.44 | 0.0879 | 0.0049 |
| R | 121 | 1185.80515325 | 0.00002416 | 25.18 | 1.71 | 0.0721 | 0.0042 |
| R | 122 | 1185.81300636 | 0.00002014 | 26.29 | 1.42 | 0.0708 | 0.0031 |
| R | 123 | 1185.82071021 | 0.00001860 | 19.63 | 1.31 | 0.0700 | 0.0041 |
| R | 124 | 1185.82837197 | 0.00003296 | 34.45 | 2.33 | 0.0715 | 0.0042 |
| R | 125 | 1185.83630935 | 0.00002124 | 25.91 | 1.50 | 0.0833 | 0.0042 |
| R | 126 | 1185.84397063 | 0.00001532 | 14.80 | 1.08 | 0.0832 | 0.0053 |
| R | 127 | 1185.85171409 | 0.00001655 | 15.13 | 1.17 | 0.0693 | 0.0046 |
| R | 128 | 1185.85945447 | 0.00001980 | 19.44 | 1.28 | 0.0776 | 0.0046 |
| R | 129 | 1185.86723090 | 0.00002003 | 21.30 | 1.41 | 0.0720 | 0.0041 |
| R | 130 | 1185.87499018 | 0.00001291 | 16.93 | 0.91 | 0.0914 | 0.0043 |
| R | 131 | 1185.88271027 | 0.00001954 | 15.77 | 1.38 | 0.0783 | 0.0059 |
| R | 132 | 1185.89048580 | 0.00002045 | 27.89 | 1.44 | 0.1051 | 0.0047 |
| R | 133 | 1185.89831131 | 0.00002719 | 26.00 | 1.92 | 0.0740 | 0.0047 |
| R | 134 | 1185.90600900 | 0.00002439 | 18.55 | 1.86 | 0.0730 | 0.0072 |
| R | 135 | 1185.91374619 | 0.00001616 | 18.83 | 1.14 | 0.0904 | 0.0047 |
| R | 136 | 1185.92153041 | 0.00001960 | 26.21 | 1.38 | 0.0956 | 0.0044 |
| R | 137 | 1185.92929557 | 0.00001914 | 17.19 | 1.35 | 0.0756 | 0.0051 |
| R | 138 | 1185.93701210 | 0.00001066 | 16.48 | 0.75 | 0.0984 | 0.0039 |
| R | 139 | 1185.94480059 | 0.00001620 | 18.37 | 1.14 | 0.0851 | 0.0046 |
| R | 140 | 1185.95258422 | 0.00001861 | 21.19 | 1.26 | 0.1052 | 0.0059 |
| R | 141 | 1185.96027741 | 0.00001746 | 18.45 | 1.23 | 0.0899 | 0.0052 |
| R | 142 | 1185.96805973 | 0.00002046 | 25.56 | 1.44 | 0.1016 | 0.0050 |
| R | 143 | 1185.97585529 | 0.00002827 | 19.90 | 2.00 | 0.0727 | 0.0063 |
| R | 144 | 1185.98351867 | 0.00001217 | 14.50 | 0.86 | 0.1018 | 0.0052 |
| R | 145 | 1185.99129801 | 0.00001223 | 16.65 | 0.86 | 0.0989 | 0.0044 |
| R | 146 | 1185.99902453 | 0.00001259 | 16.01 | 0.92 | 0.1100 | 0.0053 |
| R | 147 | 1186.00676011 | 0.00001487 | 17.50 | 1.05 | 0.1006 | 0.0052 |
| R | 148 | 1186.01449179 | 0.00001136 | 15.68 | 0.80 | 0.1002 | 0.0044 |
| R | 149 | 1186.02220786 | 0.00001804 | 24.40 | 1.27 | 0.0987 | 0.0045 |
| R | 150 | 1186.03001614 | 0.00001360 | 25.10 | 0.96 | 0.1262 | 0.0042 |
| R | 151 | 1186.03772855 | 0.00001409 | 20.62 | 1.00 | 0.0978 | 0.0041 |
| R | 152 | 1186.04545773 | 0.00001200 | 20.61 | 0.85 | 0.0981 | 0.0035 |
| R | 153 | 1186.05321004 | 0.00001675 | 22.09 | 1.18 | 0.0973 | 0.0045 |
| R | 154 | 1186.06097027 | 0.00001853 | 19.54 | 1.31 | 0.1014 | 0.0059 |
| R | 155 | 1186.06871718 | 0.00002094 | 16.79 | 1.48 | 0.0829 | 0.0063 |
| R | 156 | 1186.07642558 | 0.00001295 | 16.95 | 0.91 | 0.1148 | 0.0054 |
| R | 157 | 1186.08418307 | 0.00001604 | 21.55 | 1.17 | 0.1015 | 0.0045 |
| R | 158 | 1186.09192937 | 0.00001537 | 22.66 | 1.09 | 0.0998 | 0.0041 |
| R | 159 | 1186.09963778 | 0.00001438 | 19.65 | 1.02 | 0.0982 | 0.0044 |
| R | 160 | 1186.10738918 | 0.00001095 | 18.62 | 0.77 | 0.1240 | 0.0045 |
| R | 161 | 1186.11513853 | 0.00001351 | 19.21 | 0.95 | 0.0964 | 0.0041 |
| R | 162 | 1186.12284990 | 0.00001509 | 17.86 | 1.07 | 0.1197 | 0.0062 |
| R | 163 | 1186.13060482 | 0.00002077 | 19.44 | 1.34 | 0.0959 | 0.0060 |
| R | 164 | 1186.13832988 | 0.00001974 | 17.53 | 1.39 | 0.0935 | 0.0064 |
| R | 165 | 1186.14607679 | 0.00001656 | 17.79 | 1.17 | 0.1056 | 0.0060 |
| R | 166 | 1186.15377851 | 0.00001322 | 19.09 | 0.93 | 0.1178 | 0.0050 |
| R | 167 | 1186.16152842 | 0.00001645 | 22.74 | 1.16 | 0.1072 | 0.0047 |
| R | 168 | 1186.16925271 | 0.00001354 | 19.67 | 0.96 | 0.1257 | 0.0053 |
| R | 169 | 1186.17698789 | 0.00001497 | 21.33 | 1.02 | 0.1025 | 0.0046 |
| R | 170 | 1186.18472034 | 0.00001352 | 19.87 | 0.95 | 0.1148 | 0.0048 |
| R | 171 | 1186.19243731 | 0.00001968 | 21.83 | 1.39 | 0.1216 | 0.0067 |
| R | 172 | 1186.20020052 | 0.00001716 | 14.42 | 1.21 | 0.1203 | 0.0088 |
| R | 173 | 1186.20792414 | 0.00001046 | 16.67 | 0.74 | 0.1132 | 0.0043 |
| R | 174 | 1186.21562235 | 0.00001038 | 16.69 | 0.73 | 0.1222 | 0.0046 |
| R | 175 | 1186.22337610 | 0.00001185 | 17.74 | 0.87 | 0.1230 | 0.0050 |
| R | 176 | 1186.23110229 | 0.00000911 | 15.85 | 0.64 | 0.1137 | 0.0040 |
| R | 177 | 1186.23880472 | 0.00000990 | 16.71 | 0.70 | 0.1221 | 0.0044 |
| R | 178 | 1186.24652716 | 0.00001232 | 20.67 | 0.87 | 0.1395 | 0.0051 |
| R | 179 | 1186.25426457 | 0.00001249 | 17.60 | 0.88 | 0.1110 | 0.0048 |
| R | 180 | 1186.26199875 | 0.00000803 | 15.99 | 0.57 | 0.1299 | 0.0040 |
| R | 181 | 1186.26972423 | 0.00000980 | 16.54 | 0.69 | 0.1172 | 0.0042 |
| R | 182 | 1186.27743638 | 0.00001181 | 17.88 | 0.83 | 0.1104 | 0.0045 |
| R | 183 | 1186.28515380 | 0.00000985 | 15.40 | 0.70 | 0.1162 | 0.0045 |
| R | 184 | 1186.29287109 | 0.00000827 | 15.34 | 0.58 | 0.1216 | 0.0040 |
| R | 185 | 1186.30057909 | 0.00001579 | 18.56 | 1.11 | 0.1023 | 0.0053 |
| R | 186 | 1186.30835263 | 0.00001044 | 19.50 | 0.71 | 0.1353 | 0.0040 |
| R | 187 | 1186.31605358 | 0.00001004 | 17.41 | 0.71 | 0.1137 | 0.0040 |
| R | 188 | 1186.32377260 | 0.00001370 | 16.69 | 0.97 | 0.1114 | 0.0056 |
| R | 189 | 1186.33150299 | 0.00001384 | 15.98 | 0.98 | 0.1140 | 0.0060 |
| R | 190 | 1186.33921676 | 0.00001113 | 19.32 | 0.79 | 0.1337 | 0.0047 |
| R | 191 | 1186.34695558 | 0.00001377 | 14.67 | 0.97 | 0.1093 | 0.0063 |
| R | 192 | 1186.35465968 | 0.00001314 | 15.03 | 0.85 | 0.1272 | 0.0070 |
| R | 193 | 1186.36239364 | 0.00001202 | 16.49 | 0.85 | 0.1100 | 0.0049 |
| R | 194 | 1186.37009762 | 0.00001119 | 15.66 | 0.79 | 0.1035 | 0.0045 |
| R | 195 | 1186.37782009 | 0.00001053 | 16.00 | 0.74 | 0.1300 | 0.0052 |
| R | 196 | 1186.38553942 | 0.00000959 | 16.52 | 0.68 | 0.1297 | 0.0046 |
| R | 197 | 1186.39323692 | 0.00001149 | 16.21 | 0.81 | 0.1046 | 0.0045 |
| R | 198 | 1186.40096105 | 0.00000907 | 15.41 | 0.65 | 0.1207 | 0.0043 |
| R | 199 | 1186.40865637 | 0.00001236 | 15.75 | 0.87 | 0.0985 | 0.0047 |
| R | 200 | 1186.41643408 | 0.00001252 | 16.38 | 0.88 | 0.1122 | 0.0052 |
| R | 201 | 1186.42412871 | 0.00000721 | 17.12 | 0.51 | 0.1320 | 0.0034 |
| R | 202 | 1186.43185019 | 0.00001113 | 16.87 | 0.79 | 0.1214 | 0.0049 |
| R | 203 | 1186.43955701 | 0.00001257 | 16.23 | 0.89 | 0.1015 | 0.0048 |
| R | 204 | 1186.44725348 | 0.00001081 | 16.42 | 0.76 | 0.1134 | 0.0046 |
| R | 205 | 1186.45499854 | 0.00001037 | 16.80 | 0.73 | 0.1081 | 0.0041 |
| R | 206 | 1186.46270031 | 0.00000935 | 15.74 | 0.66 | 0.1186 | 0.0043 |
| R | 207 | 1186.47039538 | 0.00001012 | 16.15 | 0.71 | 0.1092 | 0.0042 |
| R | 208 | 1186.47808243 | 0.00000841 | 16.50 | 0.59 | 0.1169 | 0.0036 |
| R | 209 | 1186.48579442 | 0.00001380 | 18.90 | 0.95 | 0.0865 | 0.0042 |
| R | 210 | 1186.49347467 | 0.00001148 | 16.06 | 0.81 | 0.1140 | 0.0050 |
| R | 211 | 1186.50114337 | 0.00001520 | 18.00 | 1.07 | 0.0936 | 0.0048 |
| R | 212 | 1186.50885975 | 0.00001096 | 19.73 | 0.77 | 0.0973 | 0.0033 |
| R | 213 | 1186.51656248 | 0.00002158 | 25.22 | 1.52 | 0.0802 | 0.0042 |
| R | 214 | 1186.52421866 | 0.00001630 | 21.29 | 1.15 | 0.0736 | 0.0034 |
| R | 215 | 1186.53209836 | 0.00002436 | 30.00 | 1.64 | 0.0595 | 0.0029 |
| R | 216 | 1186.53978741 | 0.00002455 | 30.19 | 1.73 | 0.0532 | 0.0026 |
| R | 217 | 1186.54761244 | 0.00002038 | 24.25 | 1.44 | 0.0652 | 0.0034 |
| R | 218 | 1186.55534459 | 0.00001570 | 21.39 | 1.11 | 0.0774 | 0.0035 |
| R | 219 | 1186.56309952 | 0.00002686 | 32.36 | 1.90 | 0.0784 | 0.0040 |
| R | 220 | 1186.57078332 | 0.00002312 | 15.03 | 1.63 | 0.0783 | 0.0074 |
| R | 221 | 1186.57841159 | 0.00001755 | 14.30 | 1.24 | 0.0965 | 0.0072 |
| R | 222 | 1186.58608865 | 0.00001642 | 14.68 | 1.16 | 0.0970 | 0.0066 |
| R | 223 | 1186.59378887 | 0.00003248 | 18.51 | 2.29 | 0.0825 | 0.0089 |
| R | 224 | 1186.60144044 | 0.00001420 | 16.95 | 1.00 | 0.0872 | 0.0045 |
| R | 225 | 1186.60914619 | 0.00001355 | 18.13 | 0.96 | 0.0839 | 0.0038 |
| R | 226 | 1186.61689892 | 0.00002758 | 17.82 | 1.91 | 0.0736 | 0.0063 |
| R | 227 | 1186.62457604 | 0.00002413 | 18.86 | 1.70 | 0.0645 | 0.0050 |
| R | 228 | 1186.63230577 | 0.00003746 | 32.39 | 2.64 | 0.0630 | 0.0045 |
| R | 229 | 1186.63994363 | 0.00002034 | 23.59 | 1.44 | 0.0633 | 0.0033 |
| R | 230 | 1186.64768754 | 0.00001782 | 16.47 | 1.26 | 0.0723 | 0.0048 |
| R | 231 | 1186.65538176 | 0.00001598 | 17.90 | 1.13 | 0.0724 | 0.0040 |
| R | 232 | 1186.66311862 | 0.00001496 | 17.33 | 0.96 | 0.0812 | 0.0040 |
| R | 233 | 1186.67079875 | 0.00001785 | 19.94 | 1.26 | 0.0741 | 0.0041 |
| R | 234 | 1186.67848280 | 0.00001506 | 16.53 | 1.06 | 0.0827 | 0.0046 |
| R | 235 | 1186.68616832 | 0.00001294 | 14.66 | 0.91 | 0.0851 | 0.0046 |
| R | 236 | 1186.69394465 | 0.00003491 | 33.84 | 2.46 | 0.0466 | 0.0029 |
| R | 237 | 1186.70159502 | 0.00001226 | 15.75 | 0.87 | 0.0956 | 0.0046 |
| R | 238 | 1186.70924145 | 0.00001290 | 12.92 | 0.91 | 0.0809 | 0.0049 |
| R | 239 | 1186.71696012 | 0.00001211 | 14.38 | 0.85 | 0.0782 | 0.0040 |
| R | 240 | 1186.72463295 | 0.00001143 | 16.40 | 0.81 | 0.1059 | 0.0045 |
| R | 241 | 1186.73238632 | 0.00001766 | 22.72 | 1.25 | 0.1001 | 0.0048 |
| R | 242 | 1186.74000359 | 0.00001206 | 17.05 | 0.85 | 0.0954 | 0.0041 |
| R | 243 | 1186.74772460 | 0.00002134 | 18.50 | 1.39 | 0.0943 | 0.0061 |
| R | 244 | 1186.75542004 | 0.00001936 | 16.52 | 1.37 | 0.0868 | 0.0062 |
| R 245 | 1186.76310495 | 0.00001528 | 19.08 | 1.08 | 0.0929 | 0.0045 | |
| R 246 | 1186.77076898 | 0.00001103 | 17.70 | 0.78 | 0.1045 | 0.0040 | |
| R 247 | 1186.77847906 | 0.00001635 | 20.04 | 1.15 | 0.0892 | 0.0044 | |
| R 248 | 1186.78615560 | 0.00001735 | 15.16 | 1.23 | 0.0761 | 0.0053 | |
| R 249 | 1186.79380480 | 0.00001213 | 16.31 | 0.90 | 0.0941 | 0.0042 | |
| R 250 | 1186.80154745 | 0.00001827 | 21.70 | 1.29 | 0.0943 | 0.0049 | |
| R 251 | 1186.80914515 | 0.00001403 | 18.48 | 0.99 | 0.0682 | 0.0032 | |
| R 252 | 1186.81686652 | 0.00001005 | 16.21 | 0.71 | 0.0918 | 0.0035 | |
| R 253 | 1186.82459259 | 0.00001745 | 20.98 | 1.23 | 0.0837 | 0.0043 | |
| R 254 | 1186.83226351 | 0.00003358 | 15.25 | 2.38 | 0.0599 | 0.0081 | |
| R 255 | 1186.83991344 | 0.00001913 | 13.86 | 1.35 | 0.0799 | 0.0067 | |
| R 256 | 1186.84757925 | 0.00001986 | 16.62 | 1.40 | 0.0757 | 0.0055 | |
| R 257 | 1186.85526730 | 0.00002549 | 19.24 | 1.80 | 0.0647 | 0.0052 | |
| R 258 | 1186.86304021 | 0.00001858 | 17.50 | 1.31 | 0.0685 | 0.0044 | |
| R 259 | 1186.87060153 | 0.00002146 | 21.13 | 1.52 | 0.0599 | 0.0037 | |
| R 260 | 1186.87829548 | 0.00002413 | 18.29 | 1.75 | 0.0638 | 0.0059 | |
| R 261 | 1186.88595696 | 0.00002378 | 15.11 | 1.68 | 0.0588 | 0.0057 | |
| R 262 | 1186.89359208 | 0.00002823 | 20.25 | 1.99 | 0.0676 | 0.0058 | |
| R 263 | 1186.90149148 | 0.00006144 | 28.72 | 4.34 | 0.0415 | 0.0054 | |
| R 264 | 1186.90890727 | 0.00004101 | 12.46 | 2.89 | 0.0385 | 0.0077 | |
| R 265 | 1186.91656494 | 0.00002762 | 32.01 | 1.99 | 0.0475 | 0.0025 | |
| R 266 | 1186.92426711 | 0.00002823 | 31.80 | 2.04 | 0.0559 | 0.0030 | |
| R 267 | 1186.93238705 | 0.00004437 | 48.22 | 3.13 | 0.0400 | 0.0023 | |
| R 268 | | | | | | | Line obscured, not fitted. |
| R 269 | 1186.94760886 | 0.00004808 | 33.84 | 3.39 | 0.0250 | 0.0022 | |
| R 270 | 1186.95528590 | 0.00004018 | 40.23 | 2.84 | 0.0530 | 0.0032 | |
| R 271 | 1186.96306953 | 0.00007914 | 30.95 | 5.53 | 0.0594 | 0.0089 | |
| R 272 | 1186.97060737 | 0.00002971 | 18.22 | 2.10 | 0.0666 | 0.0066 | |
| R 273 | 1186.97830157 | 0.00003481 | 29.55 | 2.46 | 0.0752 | 0.0054 | |
| R 274 | 1186.98592563 | 0.00003686 | 16.47 | 2.60 | 0.0556 | 0.0076 | |
| R 275 | 1186.99359848 | 0.00002016 | 13.96 | 1.42 | 0.0619 | 0.0055 | |
| R 276 | 1187.00118385 | 0.00002134 | 21.08 | 1.51 | 0.0753 | 0.0047 | |
| R 277 | 1187.00894545 | 0.00008556 | 49.30 | 5.99 | 0.0695 | 0.0076 | |
| R 278 | 1187.01654953 | 0.00004536 | 17.29 | 3.20 | 0.0448 | 0.0072 | |
| R 279 | 1187.02421175 | 0.00004149 | 7.28 | 2.91 | 0.0326 | 0.0113 | |
| R 280 | 1187.03185535 | 0.00001606 | 15.82 | 1.13 | 0.0691 | 0.0043 | |
| R 281 | 1187.03950923 | 0.00003973 | 16.80 | 2.80 | 0.0282 | 0.0041 | |
| R 282 | 1187.04720940 | 0.00002218 | 12.71 | 1.57 | 0.0577 | 0.0062 | |
| R 283 | 1187.05484296 | 0.00002845 | 30.25 | 2.01 | 0.0625 | 0.0036 | |
| R 284 | 1187.06264838 | 0.00003455 | 35.25 | 2.44 | 0.0559 | 0.0033 | |
| R 285 | 1187.07028938 | 0.00005310 | 17.58 | 3.75 | 0.0391 | 0.0072 | |
| R 286 | 1187.07780484 | 0.00002974 | 13.18 | 2.10 | 0.0467 | 0.0064 | |
| R 287 | 1187.08544396 | 0.00004831 | 23.86 | 3.41 | 0.0625 | 0.0077 | |
| R 288 | 1187.09320928 | 0.00003574 | 20.44 | 2.35 | 0.0708 | 0.0074 | |
| R 289 | 1187.10082022 | 0.00004011 | 25.64 | 2.83 | 0.0559 | 0.0053 | |
| R 290 | 1187.10858367 | 0.00004108 | 27.66 | 2.90 | 0.0635 | 0.0058 | |
| R 291 | 1187.11619249 | 0.00001530 | 16.74 | 1.08 | 0.0616 | 0.0034 | |

| | | | | | | |
|---|---|---|---|---|---|---|
| R 292 | 1187.12378493 | 0.00003189 | 13.28 | 2.25 | 0.0462 | 0.0068 |
| R 293 | 1187.13147483 | 0.00001438 | 19.29 | 1.03 | 0.0435 | 0.0020 |
| R 294 | 1187.13913088 | 0.00003070 | 28.68 | 2.18 | 0.0642 | 0.0042 |
| R 295 | 1187.14680545 | 0.00002984 | 19.92 | 2.11 | 0.0536 | 0.0049 |
| R 296 | 1187.15444326 | 0.00002383 | 21.81 | 1.68 | 0.0575 | 0.0038 |
| R 297 | 1187.16210434 | 0.00002018 | 25.17 | 1.42 | 0.0679 | 0.0033 |
| R 298 | 1187.16972032 | 0.00002087 | 19.59 | 1.47 | 0.0550 | 0.0036 |
| R 299 | 1187.17736489 | 0.00002570 | 18.33 | 1.67 | 0.0467 | 0.0039 |
| R 300 | 1187.18504860 | 0.00002107 | 20.06 | 1.49 | 0.0577 | 0.0037 |
| R 301 | 1187.19269127 | 0.00002980 | 17.72 | 2.10 | 0.0616 | 0.0063 |
| R 302 | 1187.20030655 | 0.00003323 | 22.76 | 2.35 | 0.0644 | 0.0057 |
| R 303 | 1187.20803730 | 0.00003697 | 32.37 | 2.61 | 0.0673 | 0.0047 |
| R 304 | 1187.21563514 | 0.00002219 | 17.64 | 1.58 | 0.0670 | 0.0052 |
| R 305 | 1187.22326750 | 0.00002150 | 20.11 | 1.52 | 0.0578 | 0.0038 |
| R 306 | 1187.23093070 | 0.00003052 | 22.58 | 2.15 | 0.0525 | 0.0043 |
| R 307 | 1187.23869592 | 0.00003952 | 19.63 | 2.79 | 0.0652 | 0.0080 |
| R 308 | 1187.24625073 | 0.00001617 | 13.64 | 1.14 | 0.0541 | 0.0039 |
| R 309 | 1187.25391044 | 0.00001833 | 20.01 | 1.29 | 0.0579 | 0.0032 |
| R 310 | 1187.26157453 | 0.00001790 | 17.54 | 1.20 | 0.0600 | 0.0039 |
| R 311 | 1187.26925509 | 0.00001644 | 19.31 | 1.16 | 0.0500 | 0.0026 |
| R 312 | 1187.27685924 | 0.00002335 | 17.11 | 1.65 | 0.0479 | 0.0040 |
| R 313 | 1187.28453206 | 0.00001863 | 16.24 | 1.32 | 0.0548 | 0.0038 |
| R 314 | 1187.29211286 | 0.00002159 | 15.16 | 1.52 | 0.0460 | 0.0040 |
| R 315 | 1187.29985955 | 0.00004309 | 30.51 | 3.11 | 0.0374 | 0.0032 |
| R 316 | 1187.30744477 | 0.00002762 | 21.00 | 1.95 | 0.0473 | 0.0038 |
| R 317 | 1187.31516486 | 0.00001950 | 17.28 | 1.38 | 0.0495 | 0.0034 |
| R 318 | 1187.32275327 | 0.00001871 | 19.04 | 1.32 | 0.0554 | 0.0033 |
| R 319 | 1187.33038469 | 0.00002145 | 22.64 | 1.51 | 0.0610 | 0.0035 |
| R 320 | 1187.33803062 | 0.00001961 | 23.00 | 1.38 | 0.0594 | 0.0031 |
| R 321 | 1187.34569783 | 0.00001991 | 21.02 | 1.35 | 0.0556 | 0.0033 |
| R 322 | 1187.35332058 | 0.00002365 | 22.28 | 1.67 | 0.0528 | 0.0034 |
| R 323 | 1187.36103138 | 0.00002012 | 22.58 | 1.42 | 0.0495 | 0.0027 |
| R 324 | 1187.36859051 | 0.00001867 | 25.08 | 1.32 | 0.0629 | 0.0029 |
| R 325 | 1187.37627897 | 0.00001760 | 22.10 | 1.24 | 0.0505 | 0.0025 |
| R 326 | 1187.38395213 | 0.00001768 | 21.32 | 1.27 | 0.0619 | 0.0031 |
| R 327 | 1187.39158828 | 0.00002449 | 32.09 | 1.75 | 0.0523 | 0.0024 |
| R 328 | 1187.39917407 | 0.00002288 | 29.47 | 1.62 | 0.0477 | 0.0023 |
| R 329 | 1187.40671845 | 0.00002574 | 32.80 | 1.82 | 0.0439 | 0.0021 |
| R 330 | 1187.41450257 | 0.00002633 | 22.77 | 1.86 | 0.0373 | 0.0026 |
| R 331 | 1187.42222185 | 0.00004998 | 33.76 | 3.53 | 0.0471 | 0.0043 |
| R 332 | 1187.42977001 | 0.00003534 | 46.50 | 2.50 | 0.0503 | 0.0024 |
| R 333 | 1187.43723706 | 0.00003374 | 39.93 | 2.38 | 0.0425 | 0.0022 |
| R 334 | 1187.44498861 | 0.00003096 | 28.04 | 2.19 | 0.0396 | 0.0027 |
| R 335 | 1187.45268561 | 0.00001893 | 15.49 | 1.34 | 0.0389 | 0.0029 |
| R 336 | 1187.46034441 | 0.00003508 | 31.17 | 2.48 | 0.0457 | 0.0031 |
| R 337 | 1187.46803953 | 0.00003990 | 28.80 | 2.87 | 0.0387 | 0.0032 |
| R 338 | 1187.47557636 | 0.00003613 | 28.43 | 2.55 | 0.0387 | 0.0030 |

| | | | | | | | |
|---|---|---|---|---|---|---|---|
| R 339 | 1187.48322006 | 0.00003718 | 26.63 | 2.62 | 0.0370 | 0.0032 | |
| R 340 | 1187.49075223 | 0.00003124 | 29.17 | 2.21 | 0.0400 | 0.0026 | |
| R 341 | 1187.49851448 | 0.00003157 | 29.95 | 2.23 | 0.0409 | 0.0026 | |
| R 342 | 1187.50608509 | 0.00002714 | 33.49 | 1.92 | 0.0448 | 0.0022 | |
| R 343 | 1187.51381102 | 0.00003351 | 30.42 | 2.26 | 0.0391 | 0.0026 | |
| R 344 | 1187.52134618 | 0.00002902 | 33.28 | 2.05 | 0.0389 | 0.0021 | |
| R 345 | 1187.52905720 | 0.00004661 | 32.70 | 3.29 | 0.0336 | 0.0029 | |
| R 346 | 1187.53671010 | 0.00003633 | 22.03 | 2.57 | 0.0347 | 0.0035 | |
| R 347 | 1187.54430594 | 0.00003621 | 24.19 | 2.56 | 0.0265 | 0.0024 | |
| R 348 | 1187.55203277 | 0.00003358 | 37.63 | 2.29 | 0.0470 | 0.0025 | |
| R 349 | 1187.55969876 | 0.00005066 | 28.27 | 3.58 | 0.0272 | 0.0030 | |
| R 350 | 1187.56715958 | 0.00004077 | 29.04 | 2.88 | 0.0349 | 0.0030 | |
| R 351 | 1187.57480010 | 0.00003001 | 24.45 | 2.12 | 0.0360 | 0.0027 | |
| R 352 | 1187.58265594 | 0.00004076 | 38.48 | 2.88 | 0.0433 | 0.0028 | |
| R 353 | 1187.59014902 | 0.00009490 | 20.78 | 6.70 | 0.0176 | 0.0049 | |
| R 354 | 1187.59763021 | 0.00002611 | 20.40 | 1.87 | 0.0396 | 0.0031 | |
| R 355 | 1187.60541239 | 0.00004333 | 28.31 | 3.06 | 0.0356 | 0.0033 | |
| R 356 | 1187.61291834 | 0.00005625 | 14.07 | 3.97 | 0.0215 | 0.0053 | |
| R 357 | 1187.62057818 | 0.00003542 | 19.59 | 2.50 | 0.0273 | 0.0030 | |
| R 358 | 1187.62852169 | 0.00005657 | 36.80 | 3.99 | 0.0215 | 0.0020 | |
| R 359 | | | | | | | Line obscured, not fitted. |
| R 360 | 1187.64345718 | 0.00007134 | 14.23 | 5.04 | 0.0159 | 0.0049 | |
| R 361 | 1187.65120987 | 0.00003946 | 26.06 | 2.79 | 0.0355 | 0.0033 | |
| R 362 | 1187.65891274 | 0.00006469 | 21.68 | 4.57 | 0.0229 | 0.0042 | |
| R 363 | | | | | | | Line obscured, not fitted. |
| R 364 | 1187.67406893 | 0.00005100 | 25.67 | 3.62 | 0.0221 | 0.0027 | |